# The Next Generation of Metadata-Oriented Testing of Research Software

Doug Mulholland
*David R. Cheriton School of Computer Science*
*University of Waterloo*
Waterloo, Canada
dmulholland@uwaterloo.ca

Paulo Alencar
*David R. Cheriton School of Computer Science*
*University of Waterloo*
Waterloo, Canada
palencar@uwaterloo.ca

Donald Cowan
*David R. Cheriton School of Computer Science*
*University of Waterloo*
Waterloo, Canada
dcowan@uwaterloo.ca

*Abstract* – Research software refers to software development tools that accelerate discovery and simplifies access to digital infrastructures. However, although research software platforms can be built increasingly more innovative and powerful than ever before, with increasing complexity there is a greater risk of failure if unplanned for and untested program scenarios arise. As systems age and are changed by different programmers the risk of a change impacting the overall system increases. In contrast, systems that are built with less emphasis on program code and more emphasis on data that describes the application can be more readily changed and maintained by individuals who are less technically skilled but are often more familiar with the application domain. Such systems can also be tested using automatically generated advanced testing regimes.

*Keywords—Metadata, research software, open science, big data, testing.*

## I. INTRODUCTION

Research software refers to software development tools that accelerate discovery and simplifies access to digital infrastructures [1]. These so-called research software platforms are an essential tool for interdisciplinary research, in which collaboration becomes a key factor, and many examples of such platforms have emerged recently in a wide variety of domains, including health, environment and astronomy.

However, although research software platforms can be built increasingly more innovative and powerful than ever before, with increasing complexity there is a greater risk of failure if unplanned for and untested program scenarios arise [2-4]. As systems age and are changed by different programmers the risk of a change impacting the overall system increases. In contrast, systems that are built with less emphasis on program code and more emphasis on data that describes the application (i.e., metadata [5]) can be more readily changed and maintained by individuals who are less technically skilled but are often more familiar with the application domain. Such systems can also be tested using automatically generated advanced testing regimes.

This paper discusses how the descriptions of metadata-driven research software systems are being transformed into automated testing regimes that exercise and stress the systems within a systematic and reproducible framework. In this way, the paper is discussing some aspects of this transformation towards the next-generation of metadata-oriented testing of research software. For more than fifteen years members of the Computer Systems Group at the University of Waterloo's David R. Cheriton School of Computer Science (UWCSG) have been building complex operational data management systems, including research software, that are comprised of more metadata and less program code.

## II. TOWARDS THE NEXT GENEARATION OF METADATA-ORIENTED TESTING

The question is: How can metadata, that is, data that describes the application, support the creation of automated testing regimes that exercise and stress the systems within a systematic and reproducible framework? In the next section, we discuss some answers for this question

### A. Towards Metadata-Driven Systems

In traditional systems programmers write detailed programs that control user interactions and data accesses in great detail. While software development kits and various utility function packages can be parameterized, the values of parameters are usually defined within the calling functions of the program code. In a system with a greater emphasis on descriptive data (metadata), definitions are stored for at least two key aspects of how that data should be managed. In particular, how data access facilities should access the actual application data (i.e., database queries, file access methods, etc.) as well as how that data should be presented and accessed by users through a user interface are all stored in a metadata storage facility such as additional database tables. A relatively simple and generalized "kernel" of data access and user interface code can retrieve configuration and context settings from the metadata storage on demand, incorporate additional data from other datasources as needed and perform the required operations.

In systems with more detailed code there is a greater requirement for highly skilled programmers, a scarce and expensive resource, to create and maintain the system. With a greater emphasis on data that describes the application, the need for programmers is reduced. As well the likelihood of coding errors is reduced.

Administration of these metadata-driven systems is performed by editing the metadata; no program code changes are required for most changes to the system. Because the metadata is stored in a structured form, such as tables in a relational database system, it can be edited using a simple forms editing facility and these changes can be made by someone without programming experience. They are often made by staff of partner organizations that have more knowledge of the application domain and much less technological capability.

### B. Metadata-Driven Testing

UWCSG members have also made significant use of automated test suites for testing and tuning system performance. As many of the systems evolved to a web-based architecture, web-based technologies for testing were adapted to a metadata-driven architecture. One version that is frequently used is based on a php-webdriver implementation. In the metadata-driven architecture, collections of webdriver actions are stored in database tables along with the results of each execution of a suite and each action within the suite. The testing framework can also identify expected results for actions. A small and simple PHP codebase was created to retrieve test suite directives from the database, pass them to the php-webdriver interface, retrieve results, store those in the database and then repeat with the next test suite directive.

### C. Test Replay and Repeatability

Test suites, expected outputs and actual outputs can all be preserved indefinitely in database tables for future analysis. As a system is changed it is highly desirable to rerun every possible test scenario to ensure that old problems that were believed to have been resolved don't reappear and that new problems don't arise.

### D. Distributed Multi-Site Testing.

In the current version of the application development framework that is used at UWCSG, called the "Web-based Information Development Environment" ("WIDE"), data entry form fields are described with several fields, including the following:

> Entity name: (identifier suitable as an SQL database column name)
>
> Data type: one of {integer, numeric, text, checkbox, select-list, ...}
>
> Required: Yes/No
>
> Maximum width: (integer value)
>
> (several other optional fields are supported by WIDE but have been omitted for this example)

The "entity name" is used in an HTML <input name="(identifier)"> data field definition; "Data type" defines the type of data to be permitted; "Required" specifies if the form can be entered successfully without any value in this field; "Maximum width" is an optional integer value that specifies the maximum number of characters that are permitted in the data field (usually specified as the "size" attribute of an <input> data entry field.

For testing purposes the test suite facility supports several test directives, including following:

```
//       text following "//" is ignored
   open    (url)
   clear   {name=element-name}
   type    (name=element-name, text)
   click   (name=element-name)
   displayScreen   // capture a screen snapshot of the
                      application (browser) window
```

These directives enable web pages to be opened, form fields to be cleared of any existing value, text to be entered into a form field and an entity on a page to be clicked. Several other directives are supported as described in the Selenium Grid server framework at http://www.seleniumhq.org/projects/grid/.

### E. Transformation-Based Generation

With both the application and test suites defined as metadata it's possible to define a variety of transformations that use the application definition to generate tests for a wide variety of aspects of the system. For example, if a field on a data entry form is defined as accepting a numeric value in the range zero to 250, tests can be generated to attempt entering data values for zero, 250, -1, 251 and a variety of other possible values. As well non-numeric values can be attempted to verify that appropriate diagnostics are generated and system response is acceptable.

To test the data entry facility, including field value validation, for a form field named "variable1" of type integer that is a required value with up to three digits on a form with a "Submit" button named "actionSubmit", a test sequence similar to the following can be used:

```
// verify that an empty form field for "variable1" is
not permitted…

open (url for the form, possibly including
logon/password sequence)

clear "name=variable1"

click "name=actionSubmit"

displayScreen                                //
record the result screen

// verify that a value of "0" in "variable1" is
acceptable…

open (url)

clear "name=variable1"

type "name=variable1","0"

click "name=actionSubmit"

displayScreen
```

### F. Logging for Diagnosis and Additional Testing

An additional component of metadata-driven systems, and indeed many traditional systems, can further aid in system testing, maintenance and analysis. Detailed application logging can help greatly to reproduce a request or series of requests that resulted in a program error. Logs should contain enough data to allow the request sequence to be completely reproduced. In multi-user scenarios, such as web-based systems, the timestamps for log entries can be invaluable for accurately reproducing the order and timing of a sequence of requests.

In the WIDE toolkit access logs are stored in database tables and periodically archived, depending on the application usage. Logs are very helpful in identifying when an entity was changed or accessed, what userid is associated with the access and what the old and new value of the entity is.

Logging data can also be used to generate many additional tests for several purposes. From a sustainability perspective, ensuring that problems don't recur is a high priority. As well, focused system performance improvements and tuning can be performed by exercising a system with frequently used requests, as identified from an analysis of the operational logs. In addition, the test engine can, itself, be tested by using testing engine log data.

*G. Software Agents for Metadata-Based Testing*

Just as a data management system can be created with a greater emphasis on metadata and a reduced emphasis on program code, so also software agents can be defined with more emphasis on metadata. We use the name "Declarative Software Agents" to describe such an entity [6]. By using descriptive data to define the agent's input, rules and actions and then log data is recorded as the output of the agent, actions such as autonomous consistency checking between two sites become much simpler to describe and perform. In many data management scenarios repeated instances of the same data are viewed as a maintenance problem and a challenge to be avoided or overcome, but with a declarative agent together with results validation, these scenarios are transformed into opportunities to achieve improved data caching and overall system performance.

*H. Self-Testing Components*

An additional test suite directive, "checkpointDB" will be added to the test suite environment in the future to enable an entire database or some part of it to be saved in a place that can be accessed by a subsequent test operation. When the system either performs an action within a test suite or fails to do so, a comparison against the checkpointed version will be made to determine whether the desired action(s) and only the desired action(s) were performed correctly. Tests like these can be run either in response to a manual request or as the result of a timed or other autonomous decision criteria.

As systems are used, maintained and age, automated testing and detailed logging are two facilities that help to ensure that the system continues to perform to its expected standards. In conclusion, with metadata-driven testing, we believe that in the future, as more metadata is captured and used, it will be possible to automate the generation of more aspects of system testing, in particular: what needs to be tested, when it should be tested and how to test it.

## III. CONCLUSIONS AND FUTURE WORK

This paper has discussed some aspects of the next generation of metadata-oriented testing of research software, aiming at addressing some key aspects in this research direction. We believe the insights described in this paper can contribute to improve research software testing, an area that has much less attention than design and implementation.